\documentclass[aip,jcp,amsmath,amsfonts,floatfix,reprint]{revtex4-1}

\usepackage{graphicx}
\usepackage{dcolumn}
\usepackage{bm}
\usepackage[utf8]{inputenc}
\usepackage{bbold}
\usepackage{makecell}
\usepackage{array}
\usepackage{fourier}
\usepackage[dvipsnames]{xcolor}

\begin{document}

\title{Wavelet-based clustering for time-series trend detection}
\author{Vincent~Talbo}
\email{vtalbo@keley-data.com}
\author{Mehdi Haddab}
\author{Derek Aubert}
\author{Redha~Moulla}
\affiliation{Keley Data, 28 Rue du Dr Finlay, F-75015 Paris, France}

\begin{abstract}
In this paper, we introduce a method performing clustering of time-series on the basis of their trend (increasing, stagnating/decreasing, and seasonal behavior). The clustering is performed using $k$-means method on a selection of coefficients obtained by discrete wavelet transform, reducing drastically the dimensionality. The method is applied on an use case for the clustering of a 864 daily sales revenue time-series for 61 retail shops. The results are presented for different mother wavelets. The importance of each wavelet coefficient and its level is discussed thanks to a principal component analysis along with a reconstruction of the signal from the selected wavelet coefficients. 
\end{abstract}

\maketitle

\section{Introduction}
With the advent of big data and cloud computing, as well as the improvement of the power of processors, the storage of big datasets is now possible. In particular, collections of observations made chronologically are stored, called time series. Time series are now collected for numerous applications, to follow the time evolution of sales, stock prices, biomedical measurements, weather data, particles in physics, etc... 

This great amount of available time-series data urged the scientific community to find ways to analyse and mine data, focusing its attention on many purposes such as : sub-sequence matching, anomaly detection, clustering, classification, patterns identification, trend analysis, segmentation and forecasting. The main issues in the analysis rely on the huge size of the data, and their great dimensionality, making the extraction of features complicated not only for humans, but also for computers. Reviews about the state-of-the-art of data mining for time-series have been published \cite{Fu2011,Esling2012}, some of them focusing on new techniques such as deep learning \cite{Fawaz2019}, or in the field of visual analytics \cite{Ali2019}.

Amongst the purposes of time series research, clustering consists in placing data into homogeneous groups\cite{Berkhin2006,Rai2010}. Objects that have maximum similarity are put in the same group, and have minimum similarity with objects in the other groups. Thanks to its unsupervised approaches, clustering is partially tackling the problem of massive datasets, leaving the task to identify patterns to the machine. However, the high feature correlation, and the potentially high noise on time series are still obstacles to reach a proper clustering. For an extensive review of clustering of time series data, the initial work by Liao \cite{Liao2005} can be completed by other reviews and textbooks published since \cite{Laxman2006,Rani2012,Aghabozorgi2015,Maharaj2019}. Time-series clustering have various applications, such as anomaly detection, whole time clustering, sub-sequence clustering and time-point clustering, all of them considering data for different time scales. 

To reach different time scales, wavelet transforms (WT) are a viable tool. Since their introduction by Morlet \textit{et al.} in 1982\cite{Morlet1982}, WT have been performed for a wide range of applications \cite{Debnath2015,Gallegati2014,Goswami2011,Meyer1993,Young1993}. Unlike discrete Fourier transforms, this method is able to capture the variations of a signal at different scales, by decomposing the signal within a time-frequency 2D space\cite{Meyer1989,Daubechies1992,Meyer1993,Young1993,Kaiser1994,Heil2006,Vidakovic2009}. Another feature of the wavelet transform is its localization : a discontinuity in the signal is only contained in one wavelet coefficient at a given scale, while in Fourier transform the discontinuity interacts with every sine/cosine functions, thus influencing all Fourier coefficients. The multi-resolution analysis (MRA) allows a fast calculation of wavelet coefficients, and is particularly suitable for the study of time series, as a smaller number of points (a subset of the wavelet coefficients) can be used to understand the behavior of a series depending on the selected scale\cite{Mallat1989,Percival2000,Chaovalit2011}.

In 1999, Huhtala \textit{et al.} calculated wavelet coefficients on financial data sets in order to look for similarities in aligned time-series\cite{Huhtala1999}. They performed cosine similarity calculation, clustering using self-organizing map, and a comparison based on selected features in order to regroup currencies with similar long-time behavior (trend), reachable from wavelet coefficients. 

In a series of papers, Vlachos \textit{et al.} proposed to enhance the quality of clustering by the use of DWT \cite{Vlachos2003}. The first step consists in extracting the wavelet coefficients using DWT on the signal with Haar wavelet. Then, $k$-means clustering is applied only to the lowest level wavelet coefficients. The obtained centres are then doubled (to fit the dimensions) and re-used to initialise $k$-means on higher level wavelet coefficients. This operation is repeated until the centres do not move between two iterations. This results in an improvement of clustering times, as well as clustering quality. Lin \textit{et al.} generalised the use of multi-resolution analysis to improve performance of other clustering methods\cite{Lin2005,Lin2004}.

Later, D'Urso, Maharaj \textit{et al.} presented the use of wavelet variances and correlations to cluster multivariate time series, on both crisp and fuzzy clustering methods \cite{Maharaj2010,DUrso2012}. In parallel, Antoniadis \textit{et al.} proposed two methods of clustering for functional data such as time-series using wavelet transform, the first using the classical clustering tools such as $k$-means, the second clustering using a dissimilarity measure between curves\cite{Antoniadis2013}.

Wavelets have also been used for economy applications: for example, one can cite the forecasting of the car sales in the Spanish market\cite{Arino2004}, the study of business cycles synchronisation and Euro\cite{AguiarConraria2011}, or a textbook edited by Gallegati and Semmler referencing wavelet applications in economics and finance\cite{Gallegati2014}. However, few papers are actually using clustering methods on wavelet coefficients calculated from sales revenue time series. 

In this paper, we present a method aiming at computing the wavelet coefficients of a time-series, then consisting in a selection of the coefficients to isolate the general trend, and finally performing a clustering of the time-series using $k$-means method with the wavelet coefficients as input. 

Using our method to a real case study, 61 retail shops have been clustered in three groups according to the trend of their sales revenues (increasing, stagnating/decreasing, and seasonal behavior), based on a time-series of 864 daily revenues. 

In the following section, the DWT technique and the clustering method are detailed. In the third section, the results of the shops use case are presented and analysed in terms of the choice of the mother wavelet, wavelets coefficients, and reconstruction of signal. A principal component analysis (PCA) is performed in order to determine the importance of each wavelet coefficients on the clustering.  


\section{Model}

Our method to classify shops according to their sales trend (increasing, stagnating, special behavior) relies in three main steps. The first step consists in performing a standard normalisation on the daily sales revenues of our dataset. Such step is important in order to not only separate shops based on their turnover. The second and third steps of our model are the computation of wavelet coefficients, and the clustering of these coefficients by the $k$-means method. 

\subsection{Wavelet transform}

The computation of the discrete wavelet transform, in order to obtain wavelet coefficients, is based on the multi-resolution analysis (MRA) introduced by Mallat \cite{Mallat1989}. The mathematical background and details of such calculation, and in general of wavelet transform, is extensively detailed in textbooks\cite{Daubechies1992,Meyer1993,Young1993,Kaiser1994,Nason2008,Mallat2009,Nason2008}, and we just provide few insights on how wavelet coefficients are obtained. 

Wavelets are short oscillations in the time domain, vanishing quickly to zero in the positive and the negative directions, with an average value of zero, and an energy of one: in other words, a wave localised both in time and frequency domains, hence the name.

In the frame of the discrete wavelet transform (DWT), a signal $f(t)$ is projected on an orthonormal basis of wavelets $\psi_{j,k}(t)$, which are scaled and translated versions of a mother wavelet $\psi(t)$: 
\begin{equation}
\psi_{j,k}(t) = \frac{1}{\sqrt{2^j}} \psi\left(\frac{t - 2^j k}{2^j}\right), (j,k) \in \mathbb{Z}^2
\label{eq1}
\end{equation}
with $(j,k)$ the scale and translation factors respectively. The wavelet coefficients $\gamma(j,k)$ are then obtained by : 
\begin{equation}
\gamma(j,k) = \int_{-\infty}^{\infty} f(t) \psi_{j,k}^{*}(t) dt.
\label{eq2}
\end{equation}

From equation \eqref{eq2}, we can see that time and frequency spaces are sampled at discrete intervals, divided by two at each scale (dyadic sampling). However, if the translations $k$ are limited by the length of the signal, we still have to perform an infinite number of scaling operations on the signal - in order to be able to reconstruct the signal afterwards. Indeed, each wavelet acts like a band-pass filter, and between two scaling coefficients $j = J$ and $j=J+1$, the bandwidth is divided by two and frequency components are also shifted by a factor of two.  As we cannot cover the spectrum up to zero-frequency, a threshold frequency is chosen and the rest of the frequencies up to zero are covered with a scaling function $\varphi$ acting like a low pass filter\cite{Mallat1989}.
If we define the space $V_J$ as the space of functions with detail up to the scale of resolution $J$, then it also includes functions with less details, meaning that $V_{J-1}\subset V_{J}$. This means that the scaling function can be written a linear combinaison of wavelets $\psi_{j,k}$ previously used for the calculation of the wavelet coefficients at the coarsest scale. 

We now have a series of wavelets $\psi_{j,k}$ acting like a band-pass filter and a scaling function $\varphi$ acting like a low-pass filter. We can then just consider that the signal is passing through a filter bank, and even avoid the representation of wavelets themselves. This is the principle of the multi-resolution analysis (MRA)\cite{Mallat1989}, a fast algorithm to perform DWT, which is used in the frame of this work. 

We now have a series of wavelets $\psi_{j,k}$ acting like a band-pass filter and a scaling function $\varphi$ acting like a low-pass filter. We can then just consider that the signal is passing through a filter bank, and even avoid the representation of wavelets themselves. This is the principle of the multi-resolution analysis\cite{Mallat1989}, a fast algorithm to perform DWT, which is used in the frame of this work. 

\begin{figure}[ht]
\centering
\includegraphics[width=3in]{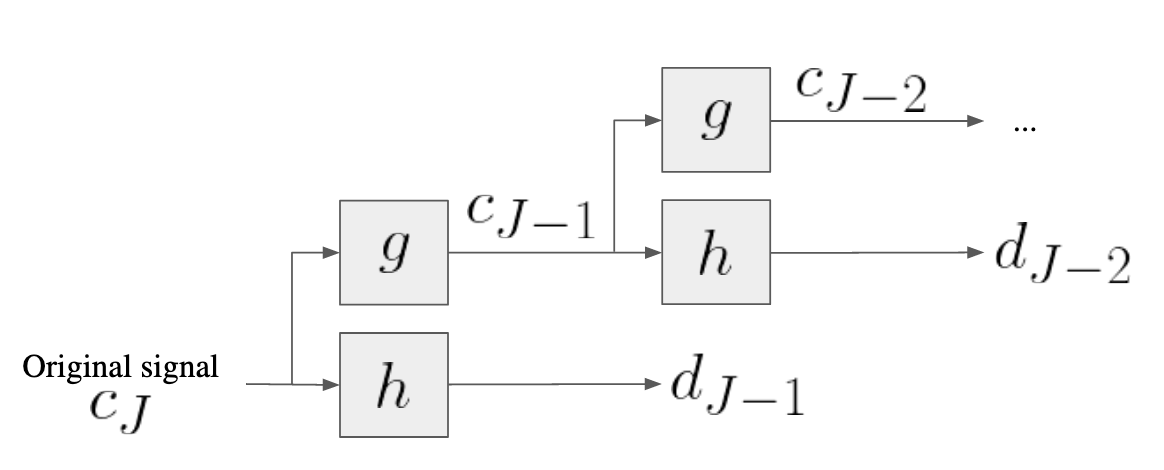}
\caption{Scheme of the multi resolution analysis algorithm using discrete wavelet transform.}
\label{fig1}
\end{figure}

The MRA consists in passing the signal (in practice, a time-series of discrete values) through a filter bank, the output of the filter bank being the wavelet coefficients. The pyramidal scheme of the MRA is given in \figurename{~\ref{fig1}}. First, we define two filters of length $M$ passing on the signal :  a high-pass filter $\{h_n\}_{(n \in \llbracket 1,M \rrbracket)}$ - equivalent to the wavelets in \eqref{eq1}-, and a low-pass filter $\{g_n\}_{(n \in \llbracket 1,M \rrbracket)}$ - equivalent to the scaling function in \eqref{eq3}. The original time-series - in our case, the daily normalised sales -  is denoted $c_{J} = (c_{J,1},c_{J,2}, ..., c_{J,N})$, where the sub-script $N$ is the number of measures, and the sub-script $J$ corresponds the maximum level of wavelet decomposition possible where at least one wavelet coefficient is uncorrupted by edge effects caused by the padding (details about padding are given later on this section). It is defined as $J = {\lfloor}{\log_2(N/(M-1))}{\rfloor}$.

The time-series is passed through the two filters $\{h_n\}$ and $\{g_n\}$, resulting in the "approximation" and "details" coefficients $c_{J-1,k}$ and $d_{J-1,k}$ respectively, using the formulas below. The size of the subset is divided by 2 compared to the original signal. Then, the approximation coefficients are passed again through the filter, and we can write the general formulas : 
\begin{align}
c_{j-1,k}= \sum_n h_n c_{j,n+2k} \quad  \text{for ${j \in \llbracket 1,J \rrbracket}$,}\label{eq1}\\
d_{j-1,k}= \sum_n g_n c_{j,n+2k} \quad \text{for ${j \in \llbracket 1,J \rrbracket}$},\label{eq2}
\end{align}
we reach iteratively the wavelet coefficients set $(c_{0},d_{0},d_{1},...,d_{J-1})$, the subscript $0$ containing the coarsest resolutions, and $J-1$ the finest. 

For non-dyadic series, and/or for a filter length higher than 2, the issue of boundaries must be addressed. For example, the reference wavelet "symlet 2" used for this work has a filter length of 4. On the left side of the series $(x_0,x_1, ...,x_N)$, the first approximation coefficients are given by $\sum_{k=0}^3 g_k x_{k-2}$. Different strategies are proposed to tackle this issue. Some of them implies a modification of the wavelet to fit the boundaries. The other methods artificially extend the boundaries by padding the series on the left (and the right for symmetry reasons). Amongst the panel of possibilities, one should mention : symmetric padding corresponds to "mirror" the values according to the boundary $(x_{-1}=x_0)$, periodic padding consists in pasting the end of the signal at the beginning of it $(x_{-1}=x_N)$, and zero-padding fills the missing values with 0s. We have chosen the latter option for the work of this paper, in order to keep the same padding for all time-series and coherently compare the wavelet coefficients for different series. 

Our wavelet coefficient calculations have been performed using the PyWavelets library for Python\cite{Lee2019}.

\subsection{Clustering wavelets}
The last step of our model consists in the clustering of time-series based on their trend. As the global trend is a long-time feature, we only select the wavelet coefficients corresponding to low frequencies, i.e., the low-level ones. Unlike Vlachos \textit{et al.}\cite{Vlachos2003} or Antoniadis \textit{et al.}\cite{Antoniadis2013}, we choose to perform a clustering on the coefficients $c_0,d_0,$ and $d_1$, without any propagation to higher level wavelet coefficients. The total number of coefficients depends on the filter length, as for the chosen mother wavelet for the wavelet decomposition, as we will see in the next section. 

The clustering method we chose for the model is $k$-means \cite{Hartigan1979}. It relies in the partitioning of $N$ observations (in our case, sets of wavelet coefficients corresponding to each time-series) into $k$ disjoint clusters $S = {S_1,S_2,...,S_k}$ with $k<=n$. The $k$-means method aims to minimise the within-cluster sum of squares, i.e. : 
\begin{equation}
\underset{\mathbf{S}}{\operatorname{arg\,min}} \sum_{i=1}^{k} \sum_{x_j \in S_i} \|x_j - \mu_i\|^2
\label{eq8}
\end{equation}
with $(x_1,...,x_N)$ the set of observations, and the $\mu_i$ the centroids of the clusters $S_i$. 

The algorithm, sometimes referred to as Lloyd's algorithm, consists in three steps: first, an initialisation process in order to choose $k$ centroids (read below). Then, the algorithm loops between the two next steps : each observation are assigned to the closest centroid, and finally, centroids are updated according to the new assignments of observations. The loop ends when the distances between the old and the new centroids is smaller than a given threshold. 

The initiation step is actually of strong importance. The standard initialisation method consists in selecting $k$ random observations and set them as centroids. The loop in the $k$-means algorithm is thus highly dependent on seeding and one has to perform the clustering many times to trust the final partitioning. The other initialisation technique is the $k$-means++ algorithm, which is used for this study : first, an observation is randomly picked as the first centroid. Then, a new centroid is chosen using a weighted probability distribution proportional to the distance between the points and their closest centroid. This step is repeated until the $k$ centroids have been chosen. This theoretically ensures that centroids are distant enough to reduce the number of steps in the $k$-means algorithm, thus improving the computational cost\cite{Arthur2007}. 
In this work, the $k$-means clustering with $k$-means++ initialisation is performed by the scikit-learn library\cite{Pedregosa2011}.


\section{Results and discussion}
\subsection{Dataset}
\begin{figure}[ht]
\centering
\includegraphics[width=3in]{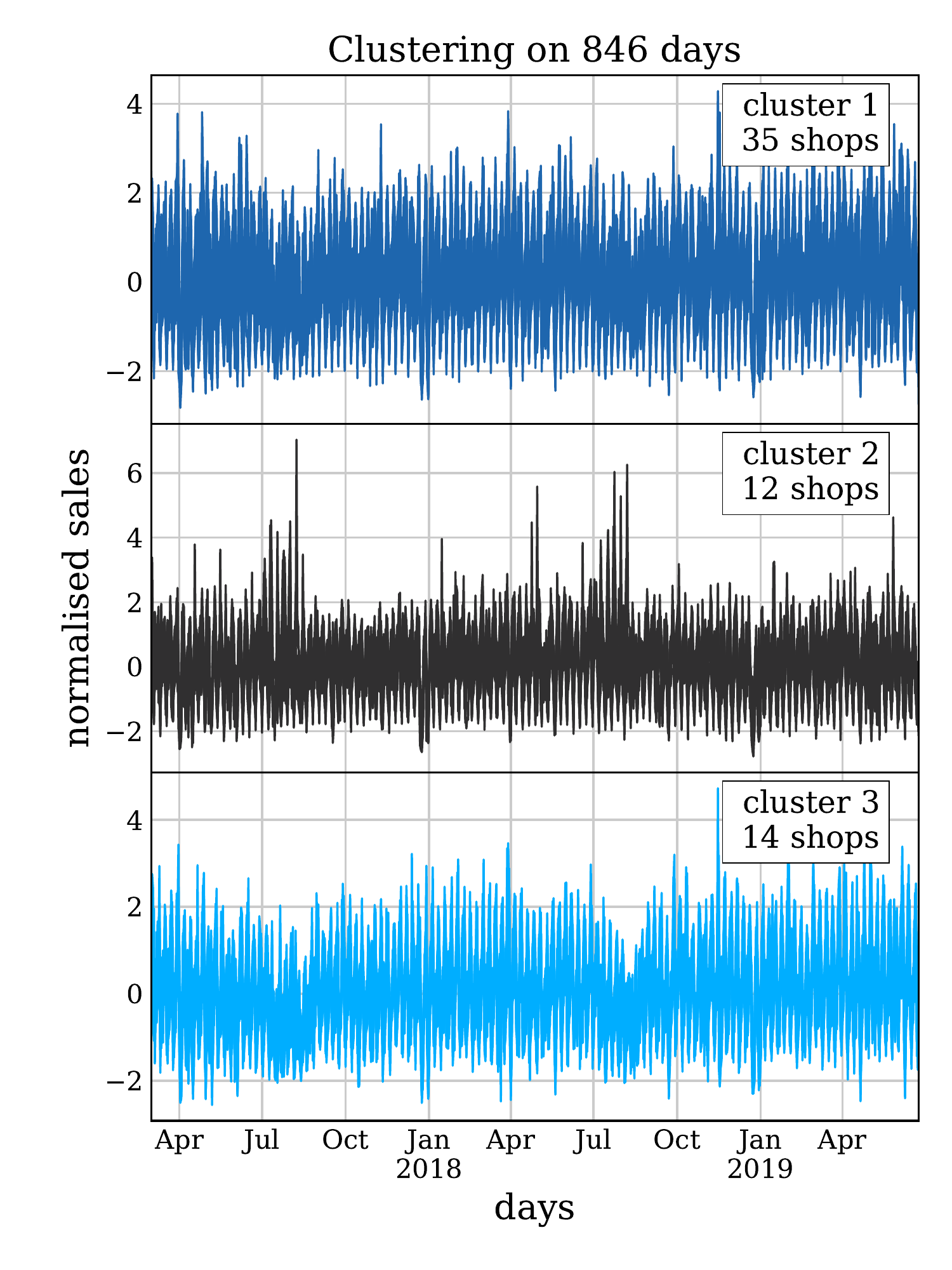}
\caption{Clustering obtained on the 846 points corresponding to daily revenues.}
\label{fig2}
\end{figure}

The dataset contains the daily sales revenues of 61 shops, from a time-period going from March 3\textsuperscript{rd}, 2017 to June 24\textsuperscript{th}, 2019 (846 days). On \figurename{~\ref{fig2}}, the result of the clustering on the 846 points shows that no trend can be extracted from the raw data. Indeed, clustering with a number of features much higher than the number of samples (61) is not really relevant, hence the necessity to reduce features dimensionality. 

\subsection{Wavelet coefficients}
\begin{figure}[ht]
\centering
\includegraphics[width=3.2in]{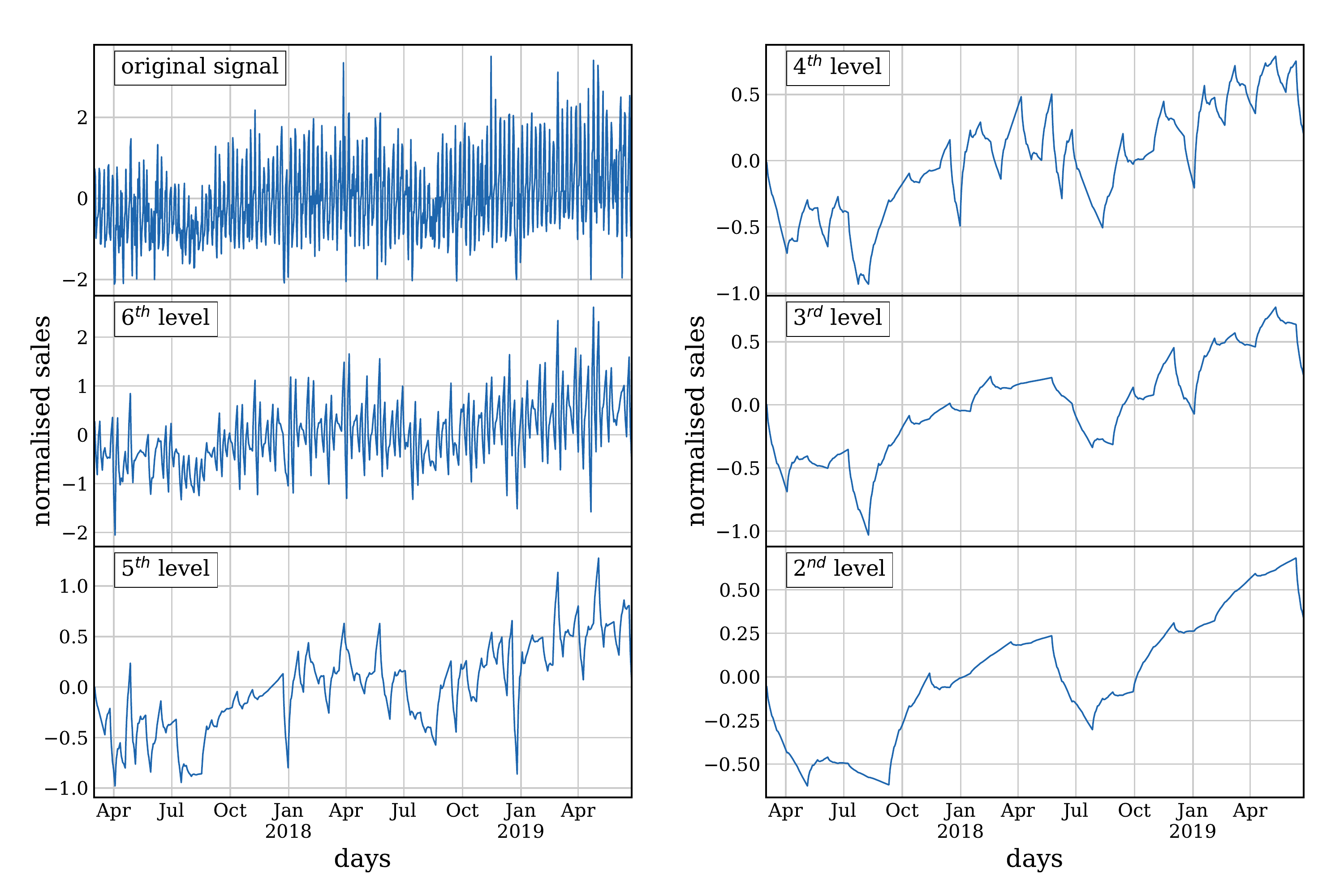}
\caption{Original time-series of the normalised sales for shop 51, and reconstructed signal from different levels of wavelet coefficients. "3\textsuperscript{rd} level" means that only ($c_0, d_0$, $d_1$, $d_2$, $d_3$) coefficients are considered, and the higher level ones are set to 0. Wavelet coefficients have been calculated by symlet 2 mother wavelet.}
\label{fig3}
\end{figure}

A multi-resolution analysis, based on DWT, was performed through the use of 15 mother wavelets. As an example, we have plotted in \figurename{~\ref{fig3}} the original signal, and the same signal reconstructed from wavelet coefficients by inverse DWT, at different scales, i.e., considering some levels of wavelets. The multi-resolution feature of the wavelet decomposition is clearly demonstrated on this figure. While the trend of the revenue sales  - increasing - is hard to extract from the original signal, it can be easily seen from the reconstruction, with up to five levels of wavelet coefficients involved. 

The list of the chosen wavelet is given in \tablename{~\ref{tab:tab1}}, with their associated filter length and the number of wavelet coefficients kept for the clustering (i.e. the coefficients $c_0, d_0$ and $d_1$). Among the wavelet library of the PyWavelet software, we have chosen the ones with the lowest filter length : a low filter length allows to reach coarser resolutions, thus revealing long-time trends, as seen in \figurename{~\ref{fig3}}. The maximum length of the wavelet coefficients series is 40, for a filter length of 6, drastically dropping the feature dimensionality by 95~\% compared to the original 846 days.  

\begin{figure}[ht]
\centering
\includegraphics[width=3in]{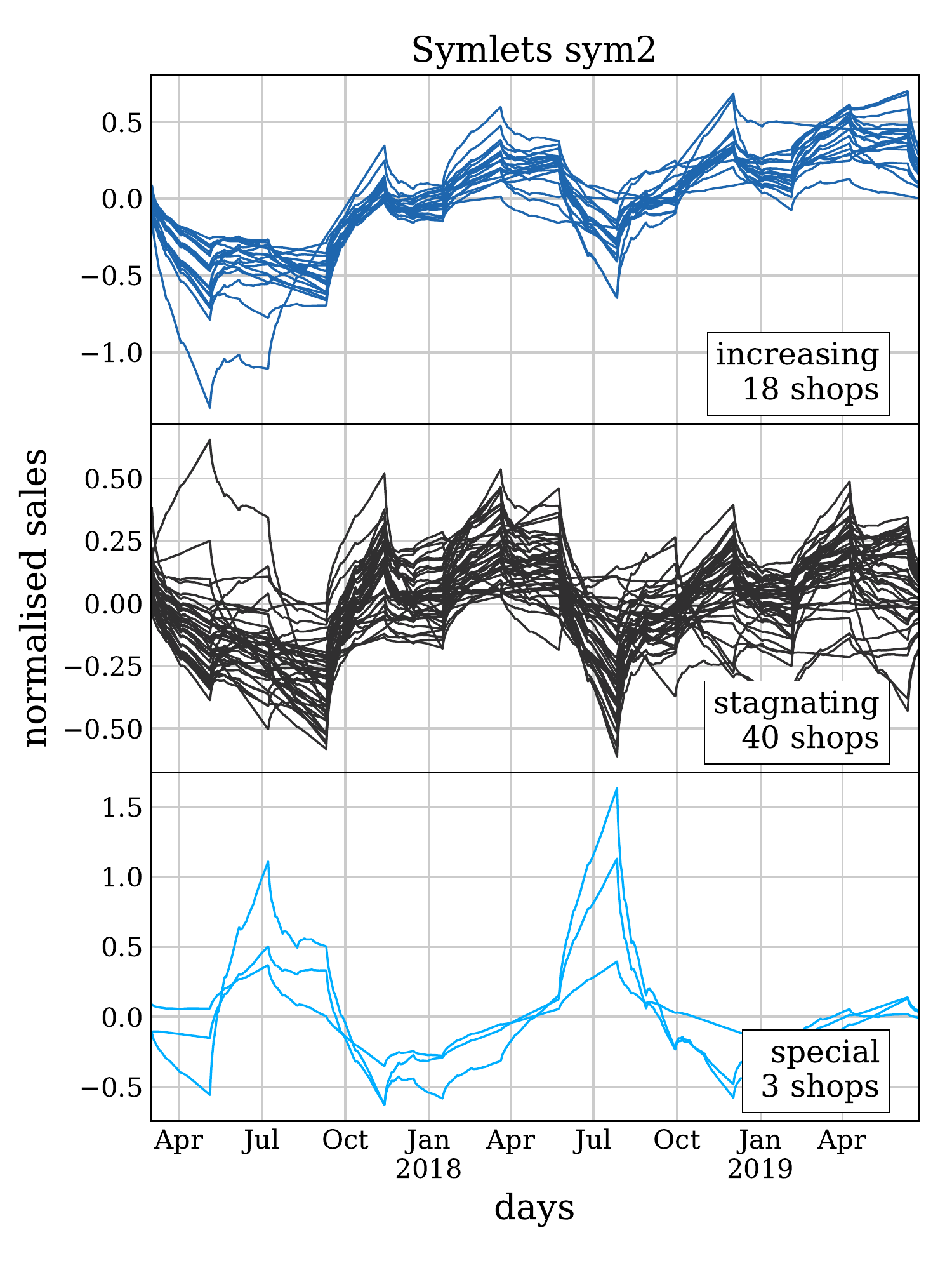}
\caption{Daily normalised sales reconstructed by inverse DWT from the 21 "symlet 2" coefficients for the 61 shops. Shops are partitioned with respect to the cluster to which they have been assigned.}
\label{fig4}
\end{figure}

\begin{table}[ht]
\begin{tabular}{ | >{\centering}m{5cm} |>{\centering}m{1.5cm} | >{\centering\arraybackslash}m{1.5cm}| } 
\hline
wavelet & filter length & $(c_0,d_0,d_1)$ length \\
\hline
\hline
Haar, Daubechies (db)  1, Biorthogonal (bior) 1, Reverse~Biorthogonal~ (rbio)~1.1 & 2 & 14 \\ 
\hline
bior3.1, db2, rbio3.1, Symlets~(sym)~2, bior1.3 & 4  & 21\\ 
\hline
bior2.2, Coiflets~(coif)~1, db3, rbio1.3, rbio2.2, sym3 & 6 & 40\\ 
\hline
\end{tabular}
\caption{List of wavelets used in this work, the length of their filter and the number of wavelet coefficients used for clustering.}
\label{tab:tab1}
\end{table}

\subsection{Clustering}

\begin{figure*}[ht]
\centering
\includegraphics[width=7in]{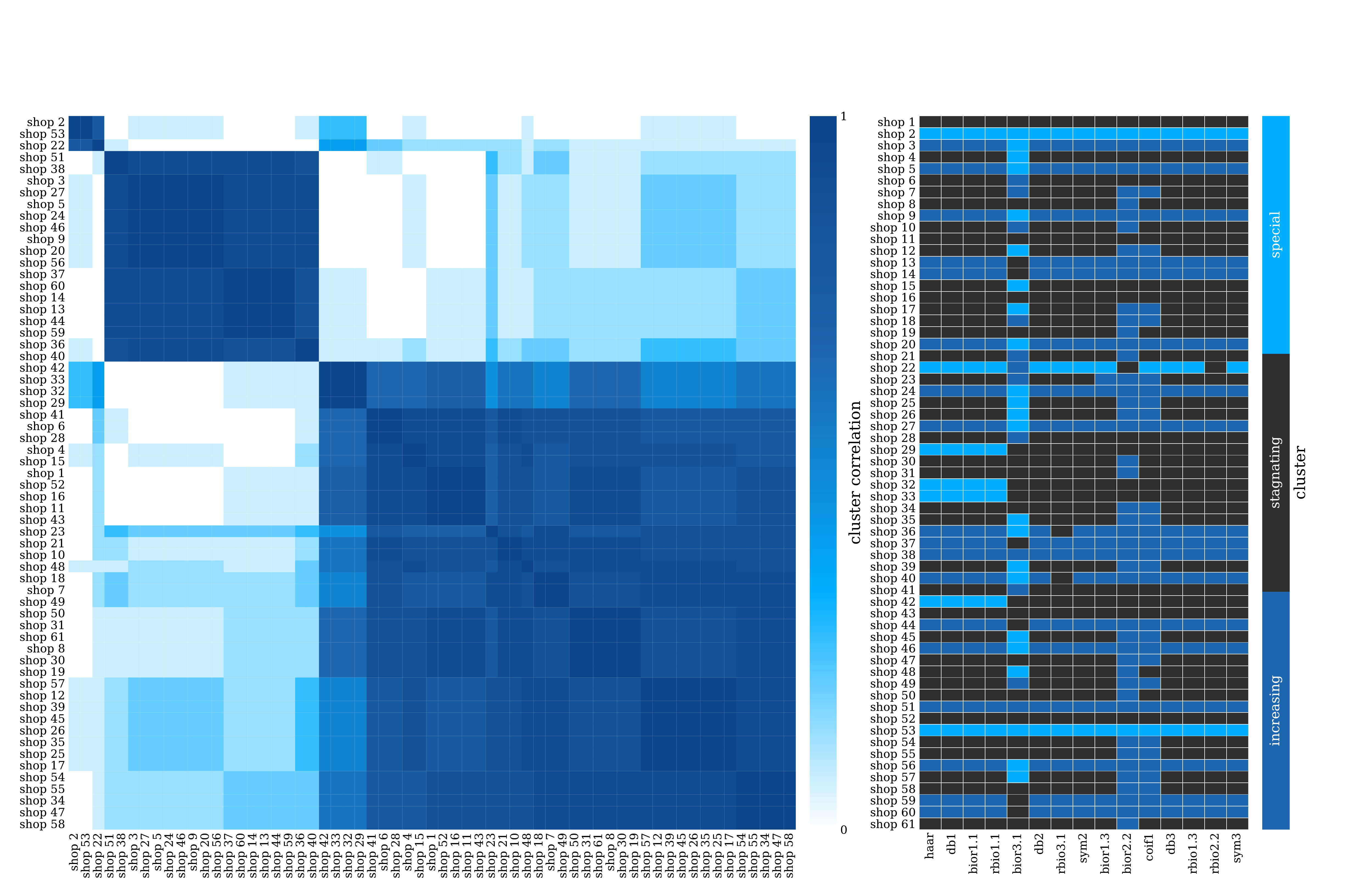}
\caption{(left)Heatmap of the correlation matrix between shops : if shops are always in the same cluster, then correlation equals 1, always in different clusters, then correlation equals 0. (right) Distribution of shops in clusters depending on the wavelet. Shops 51, 1, and 2 have been fixed to define clusters "increasing", "stagnating" and "special" respectively. }
\label{fig5}
\end{figure*}

For each of the 15 mother wavelet, a clustering using $k$-means is performed on the 14/21/40 $(c_0,d_0,d_1)$ wavelet coefficients to group the 61 shops in 3 clusters. In \figurename{~\ref{fig4}}, we see the result of the clustering by showing the reconstructed signal by inverse DWT from only $(c_0,d_0,d_1)$ coefficients (all the others are set to 0, equivalent to the last figure in \figurename{~\ref{fig3}}) obtained from the symlet 2 mother wavelet. Each of the 61 shops are plotted in the frame assigned to their cluster.

Even if the wavelet shape affects the signal, it is clear that the clustering has well separated the shops according to their sales trend, increasing, stagnating and seasonal behavior (i.e. with higher incomes in the summer, unlike the other shops). 

\subsection{Correlation matrix}

Each mother wavelet has its own properties, such as its (bi)orthogonality, its symmetry, its shape, so the clustering on the coefficients may have differed from one wavelet to another. In order to verify the stability of clustering with these different wavelets, we show on the left hand side of \figurename{~\ref{fig5}} the correlation matrix for the shops : if shops are always in the same cluster, regardless of the mother wavelet, then their correlation is 1, if they are always in different clusters, their correlation is 0. We distinguish 3 stable clusters, meaning that the choice of the wavelet only marginally affects the clusters. The figure on the right hand side in \figurename{~\ref{fig5}} shows that, if we set the clustering names "increasing", "stagnating" and "special" of shops 51, 1 and 2 respectively (because they are in different clusters for each mother wavelet), most of the wavelet keep the shops in the same clusters, with the noticeable exception of wavelet bior3.1, where the classification seems more chaotic, and bior2.2, for which more shops are included in the "increasing" group. Those two wavelets have actually a very peculiar form and do not really show the trend of sales : these features are out of the scope of this paper. Also, we notice that the 4 wavelets with a filter length of 2 tend to classify shops 29, 32, 33 and 42 as "seasonal" while other wavelets put them in the "stagnating" cluster. The wavelet coefficients associated to larger filter detect more details as more coefficients are involved in the clustering, explaining the difference of clustering separation. 

\subsection{Principal Component Analysis}

\begin{figure}[ht]
\centering
\includegraphics[width=3in]{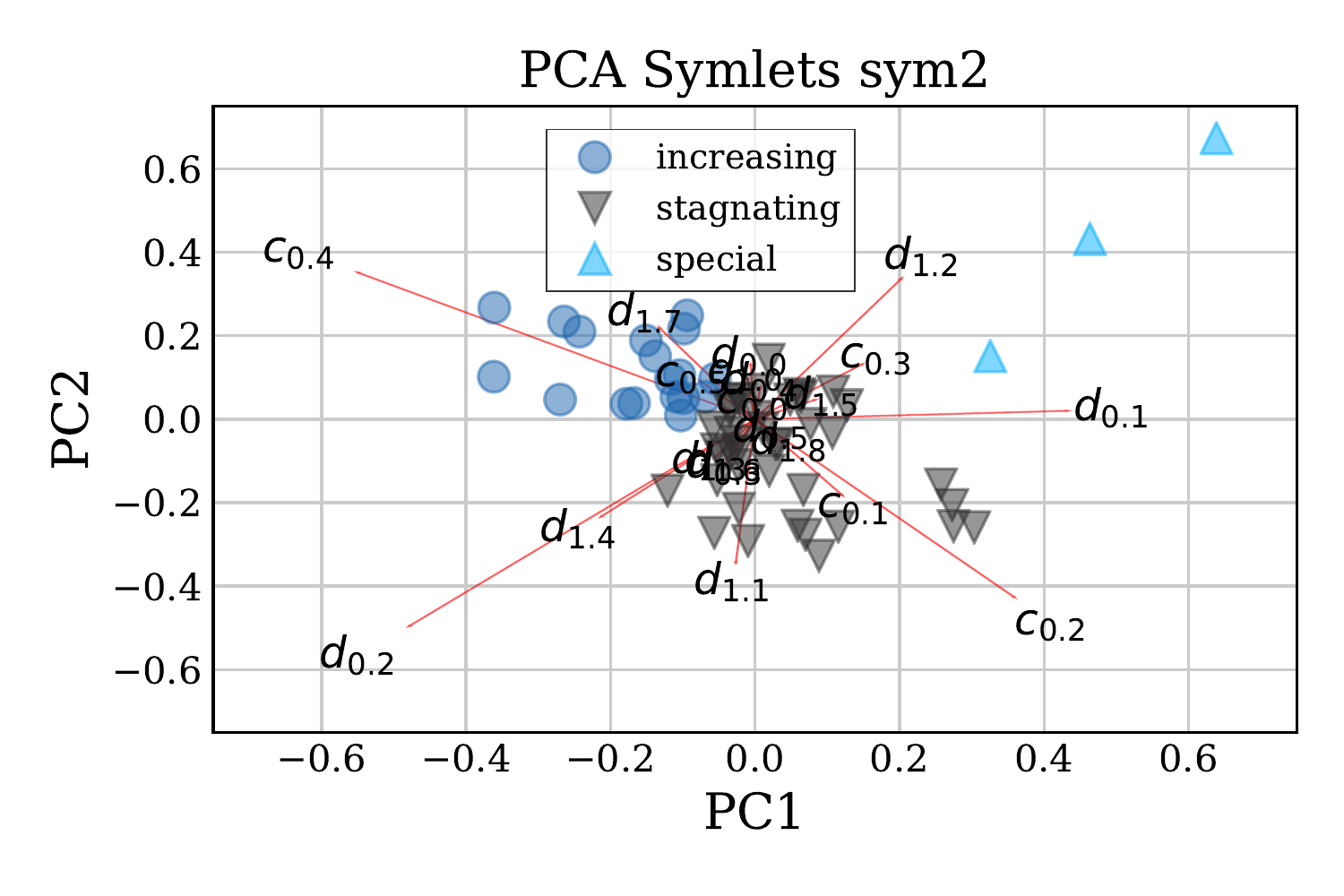}
\caption{Bi-plot of the shops wavelet coefficients, and wavelet coefficient axes, projected on the two first principal component axes obtained by a principal component analysis. Wavelet coefficients were calculated with Symlet 2 mother wavelet.}
\label{fig6}
\end{figure}

\begin{figure}[ht]
\centering
\includegraphics[width=3in]{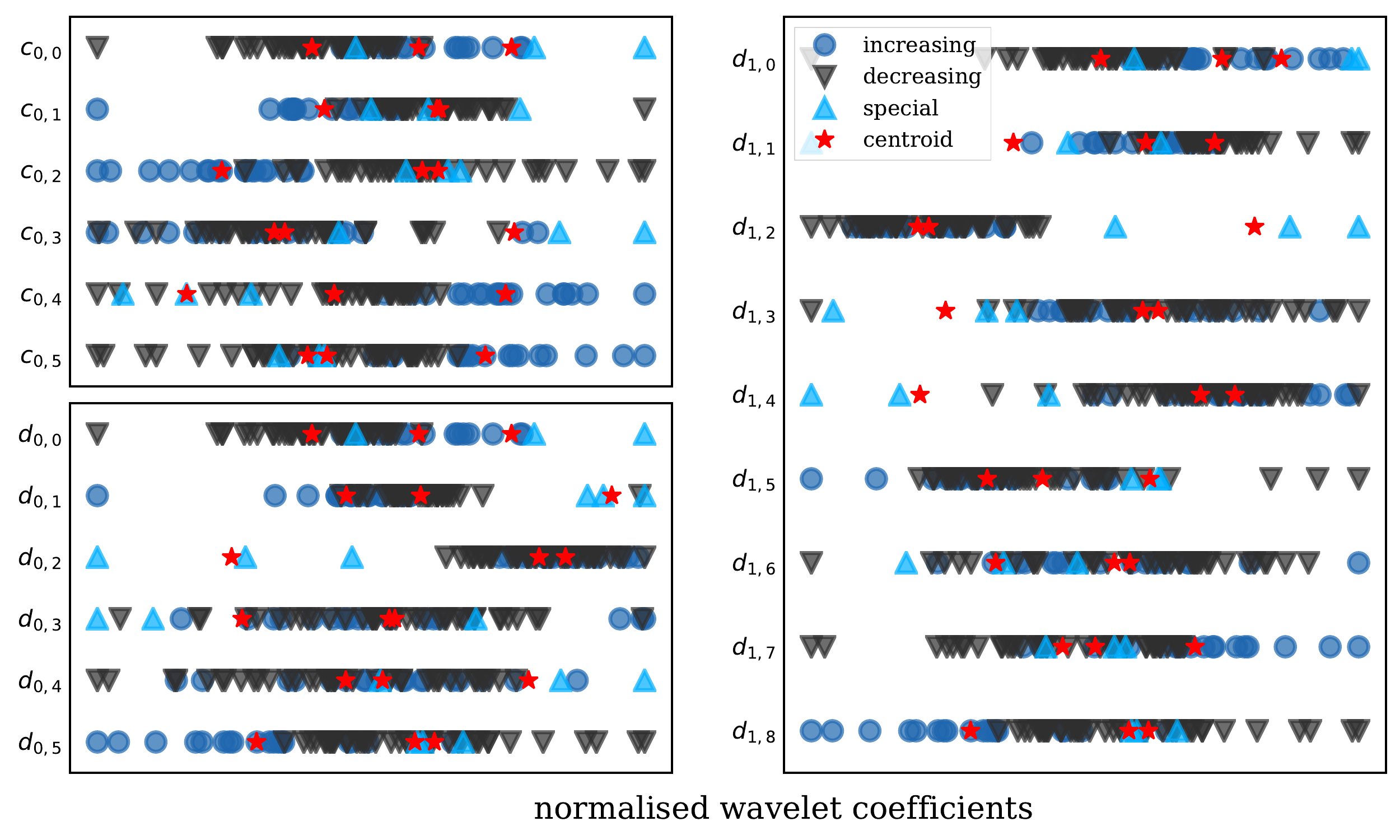}
\caption{Normalised wavelet coefficients obtained by Symlet 2 mother wavelet, for each shop. Circle, downward and upward triangles represent shops in the increasing, decreasing and special clusters respectively, and the stars indicate the centroid coordinate of the clusters.}
\label{fig7}
\end{figure}

We now have a tool, stable enough, to classify shops based on their sales trend over long periods, but we also need to localise the most important time periods for the clustering, and the economic health of shops. On this purpose, we performed a principal component analysis on the $61 \times 21$ wavelet coefficients  $(c_0,d_0,d_1)$ obtained by the Symlet 2 mother wavelet, and project the shops and the wavelet coefficient axes on the first and second principal components, shown in \figurename{~\ref{fig6}}. Also, each shop normalised wavelet coefficients are shown in \figurename{~\ref{fig7}}. With these two figures, we can see that the wavelet coefficients $c_{0,2}, c_{0,4}, d_{1,7}$ seem to separate well the clusters "increasing" and "stagnating", while the $d_{1,2}, d_{0,2}, d_{1,4} $ coefficient separate the clusters "special" from the others. Other coefficients seem to have a lower importance in the cluster decision. 

\subsection{single-coefficient signal reconstruction}
\begin{figure}[!h]
\centering
\includegraphics[width=3.2in]{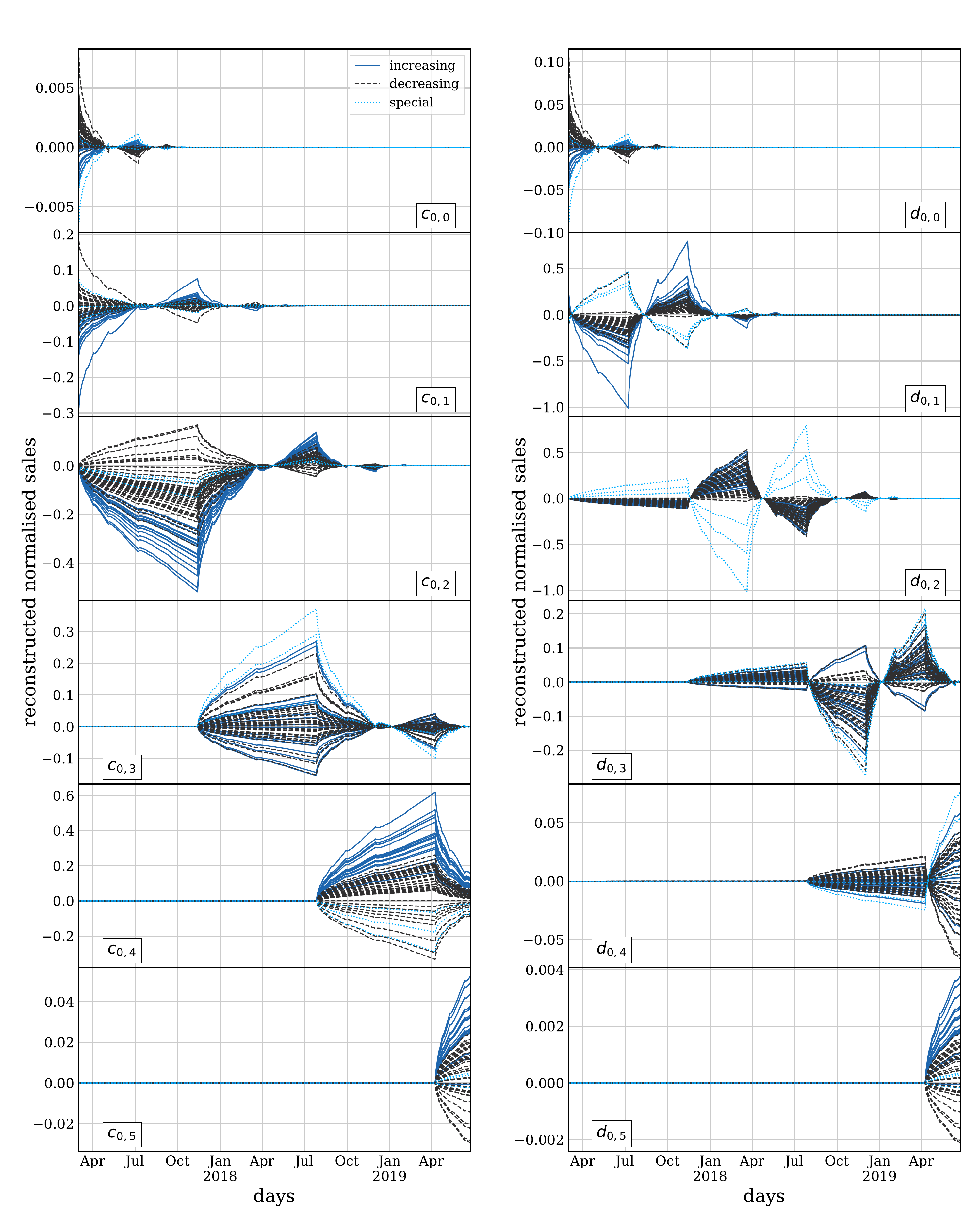}
\caption{Signal reconstruction by inverse discrete wavelet transform, from only one wavelet coefficient in $(c_0,d_0)$. Wavelet coefficients were calculated with Symlet 2 mother wavelet.}
\label{fig8}
\end{figure}
\begin{figure}[!h]
\centering
\includegraphics[width=3.2in]{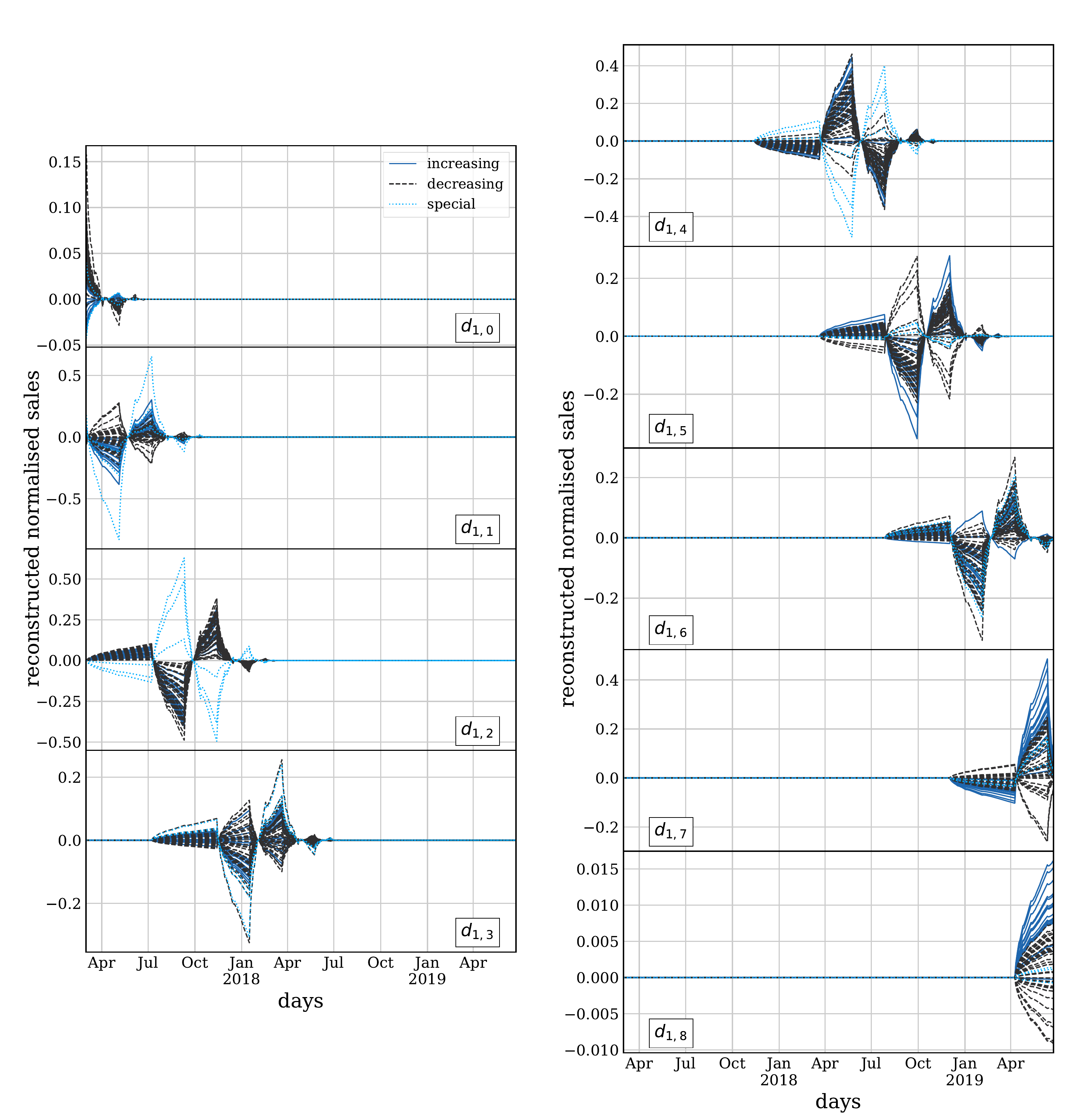}
\caption{Signal reconstruction by inverse discrete wavelet transform, from only one wavelet coefficient in $(d_1)$. Wavelet coefficients were calculated with Symlet 2 mother wavelet.}
\label{fig9}
\end{figure}

In \figurename{~\ref{fig8}} and \figurename{~\ref{fig9}}, the normalised sales revenues have been reconstructed with inverse discrete wavelet transform, with only one coefficient, all the others have been set to 0. In this representation, we are able to link a wavelet coefficient to a given period of time. For example, the coefficients $c_{0,2}$  which separates "increasing" and "stagnating", and $d_{0,2}$, separating "special" cluster from the others, correspond to a period from April 2017 to October 2018. 
$c_{0,4}$, also separating "increasing" and "stagnating", contains the sales information from August 2018 to April 2019. The reconstruction from the $d_1$ helps identifying shorter time periods for which the clustering of the trend is important. The $d_{1,7}$ coefficient separates well "increasing" from "stagnating", and span from July 2018 to April 2019. The separation between the "special" and other clusters is clearly noticeable with $d_{1,2}$, which confirms that it is linked to the seasonality: "special" shops have positive normalised sales in the summer, and negative after, while the others shops have negative normalised sales and positive after. A look at higher coefficients could show the clustering at shorter period of times, even though the clustering has been done only on coarse level.

\section{Conclusion}
We have developed a method, based on the discrete wavelet transform method and $k$-means clustering technique, to separate the trends of a long time-series. We presented an use case of a time-series of 846 daily sales revenues for 61 shops. The method consists in, after normalising the original time-series, to perform a multi-resolution analysis by discrete wavelet transform of the signal, keeping only the lowest resolution coefficients, and use the $k$-means method to group shops with similar behavior together. By doing so, the feature dimensionality is drastically reduced and allows a better clustering, which separates well the different trends. We have shown that the choice of the wavelet has low influence on the clustering, except for the reverse biorthogonal ones. The principal component analysis and the visualisation of coefficients allow us to to recognise the most important wavelet coefficients for the clustering. This analysis, along with the reconstruction of the signal based on only one given wavelet coefficient, helps identifying the period of time on which the clustering is more important, and extracts the behavior of the time-series on this period. 

\textit{Acknowledgements - }The authors would like to thank Oussama Raboun for his fruitful comments. 

\section*{Bibliography}
\bibliography{ondelettes_library}

\end{document}